\documentstyle[psfig,a4wide]{article}
\begin{document}
\begin{center}
\noindent {\Large Infinite-cluster geometry in central-force networks}
\end{center}

\begin{center}
{\it C. Moukarzel\footnote{\noindent Permanent address: Instituto de F\'\i sica,
Universidade Federal Fluminense, Niteroi RJ, Brazil. \\ \noindent  e-mail: {\bf
cristian@if.uff.br}}},\\
HLRZ/Forschungszentrum J\"ulich, D-52425 J\"ulich, Germany.\\
\end{center}

\begin{center} 
{\it P.M. Duxbury},\\
Dept. of Physics/Astronomy and Cntr. for Fundamental Materials Research,
Michigan State University, East Lansing, MI 48824-1116.\\
\end{center}

\begin{center}
and \\
{\it P.L. Leath},\\
Dept. of Physics/Astronomy, Rutgers University, Piscataway, NY 08854.\\
\end{center}

\vspace{1.0cm}

\noindent {\bf Abstract}

We show that the infinite percolating cluster (with density
$P_{\infty}$) of central-force networks is composed of: a fractal
stress-bearing backbone ($P_B$) and; rigid but unstressed ``dangling
ends'' which occupy a finite volume-fraction of the lattice ($P_D$).
Near the rigidity threshold $p_*$, there is then a {\it first-order
transition} in $P_{\infty}=P_D + P_B$, while $P_B$, is {\it
 second-order} with exponent $\beta'$ A new mean field theory shows
$\beta_{mf}'=1/2$, while simulations of triangular lattices give
$\beta_t'=0.255\pm 0.03$. \\

\noindent {\it Pacs numbers} 61.43Bn, 46.30.Cn, 05.70.Fh

\newpage

The connectivity-percolation geometry has become a paradigm in the study
of disordered systems[1], especially close to the percolation critical
point.  However, in many problems involving mechanical properties, such
 as the mechanical properties of granular media[2], glasses[3] and
gels[4], the connectivity-percolation geometry does not apply.  In
systems such as these, in which {\it central forces} are of primary
importance, the infinite cluster must be {\it multiply connected} in
order to transmit stress and hence to support any mechanical property.
Here we present a comprehensive analysis of the geometry of the infinite
cluster in this ``rigidity percolation''  problem.
\begin{figure}[] \vbox{ 
\centerline{ 
{\bf a} \psfig{figure=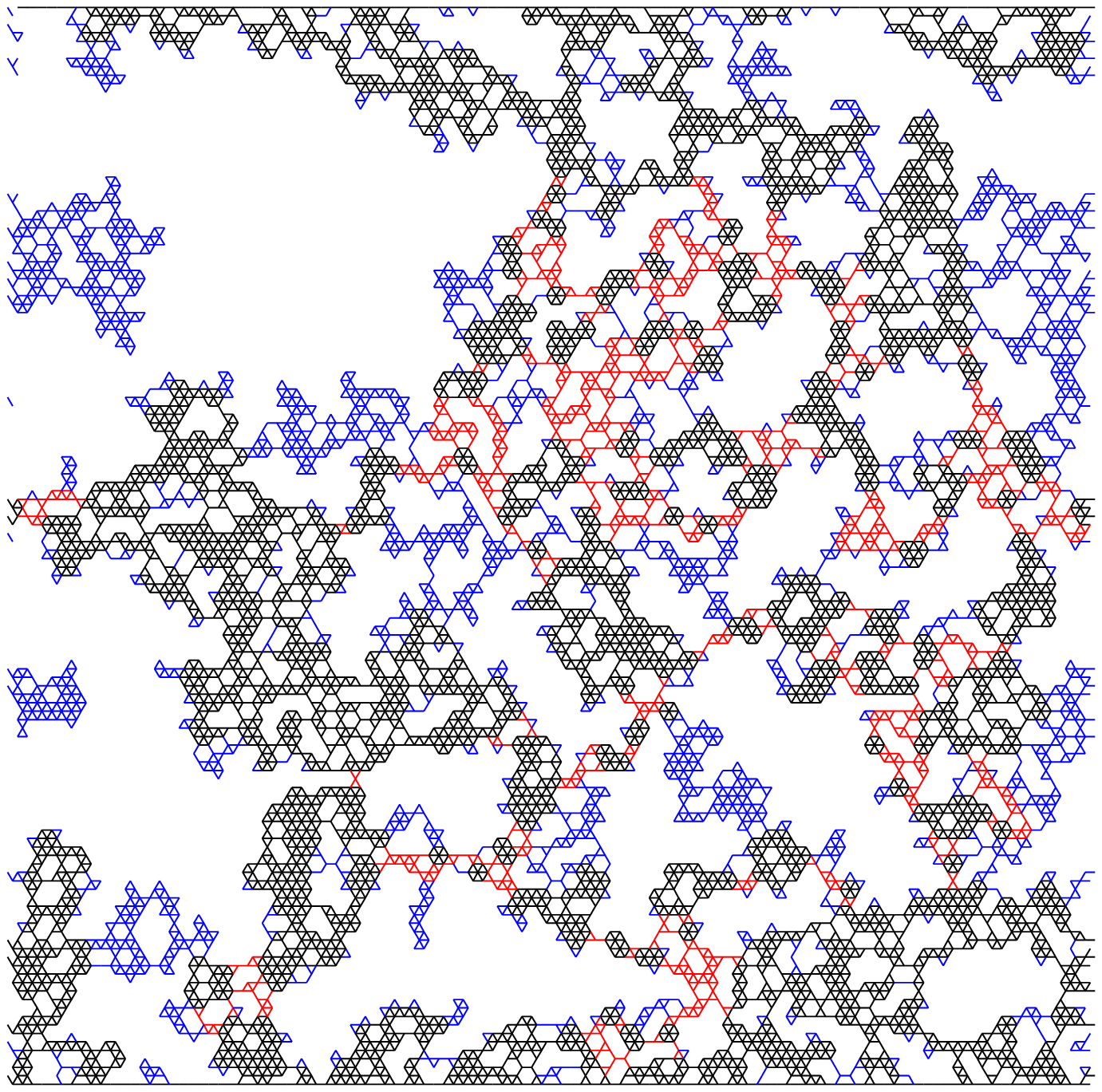,width=7cm,angle=000} \hfill 
{\bf b} \psfig{figure=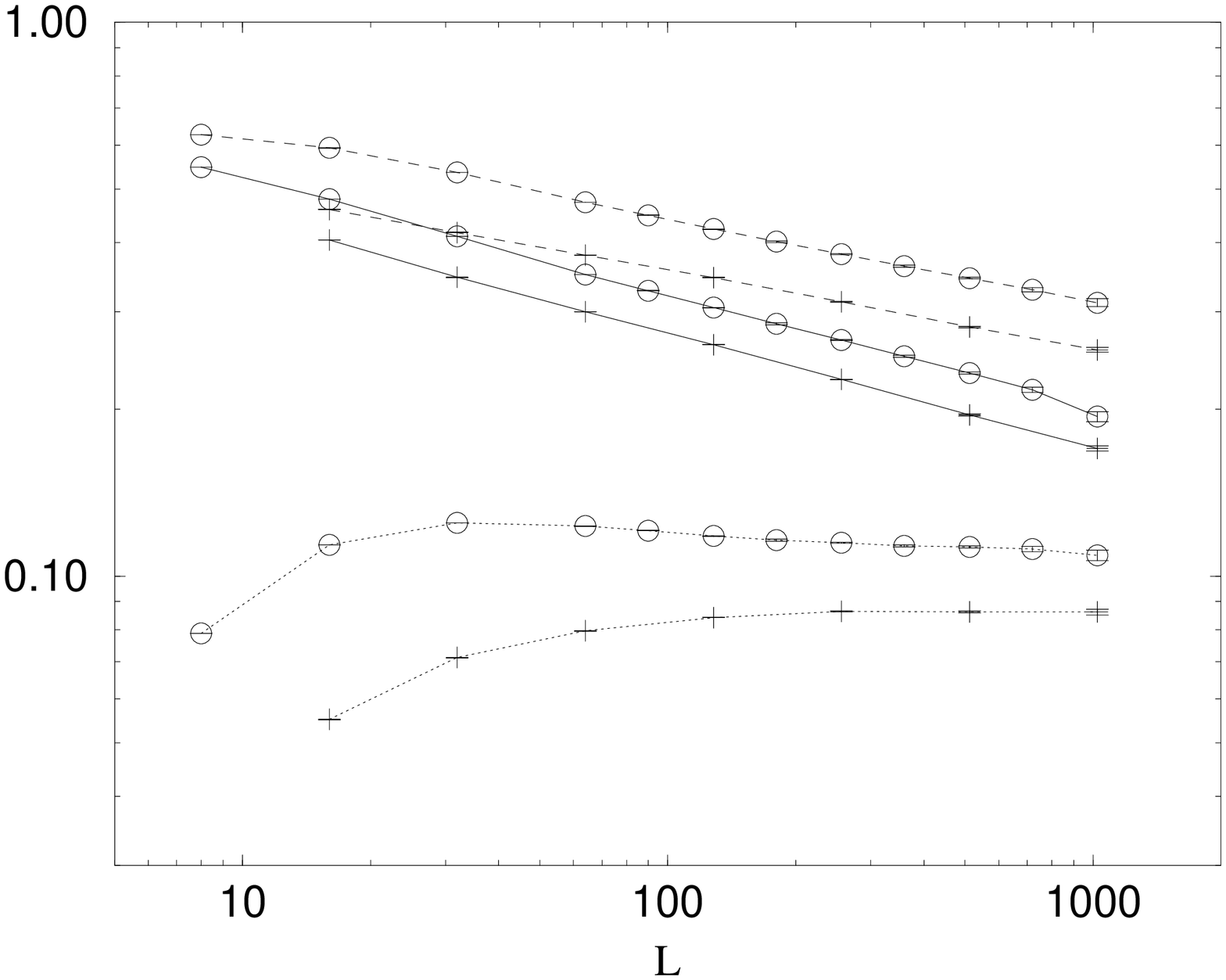,width=7cm,angle=270} 
}
\centerline{}
\caption{ 
The geometry of the infinite rigid cluster. {\bf a)}  An infinite rigid
cluster at the rigidity threshold of a site-diluted triangular lattice
($p_*=0.69707$ for this configuration). Backbone (stressed) bonds are
black, dangling ends (rigid but unstressed bonds) are blue and cutting
bonds (if they are removed the backbone collapses) are red.  {\bf b)}
$P_{\infty}$ (long dashes); $P_B$ (solid lines) and; $P_D$ (short
dashes) for site- (plusses) and bond- (circles) diluted triangular
lattices.  The calculations were done at the percolation point for each
lattice size and are averaged over roughly $10^8/L^2$ configurations.
}
\label{fig:one}
} \end{figure}
Early work on the stress-bearing paths in central-force systems relied
on direct solution to the force equations for model systems[5-8], such
as lattices of Hookian springs.  More recently, a more accurate method
which finds the backbone[9] and infinite-cluster[10] geometry directly
has been developed.  There has also been progress in the development of
mean-field theories, both continuum[11] and using Cayley trees, and we
report on the latter here.  Using a combination of these new techniques
a novel picture of rigidity percolation has emerged.  In particular, we
show that:
\begin{itemize}
\item The infinite cluster is composed of a stress-carrying backbone and
rigid (but unstressed) ``dangling ends''(see Fig. 1a).  The backbone is
{\it fractal} and this is the reason that the {\it elastic constants
undergo a second-order transition} on approach to the rigidity
 transition[5-9].
\item The ``dangling ends'' occupy a finite volume fraction of the
lattice and for this reason there is a {\it first-order transition in
$P_{\infty}$} at $p_*$
\item Due to the fractal backbone, there is a length which diverges with
a non-trivial exponent, $\nu$, at the rigidity threshold.  This exponent
is different than the connectivity-percolation correlation-length
exponent.
\item The number of ``red'' or cutting bonds scales with exponent
$1/\nu$ at the  rigidity threshold.
\end{itemize}

Consider a triangular lattice composed of nodes connected by Hooke
springs.  Consider attaching rigid ``bus-bars'' on two sides of the
lattice, and then removing $1-p$ of the sites (or bonds) from the
lattice.  Upon application of an external stress (e.g. a tensile stress)
to the lattice, we find results such as that presented in Fig. 1a.  In
this figure, the black bonds carry stress, while the ``blue'' bonds are
rigid but do not carry stress. The red bonds carry stress and are
critical in the sense that if they are removed, the lattice can no
longer support the applied stress (they are ``cutting'' bonds)[12].  The
configuration in this figure is exactly at the percolation point.  This
figure was found using a new ``graph-theory'' method[13] which enables
us to find the rigidity-percolation geometry exactly and which has been
 described elsewhere[9,10].  We plot the volume fraction of backbone,
dangling-end and infinite-cluster bonds as a function of sample size in
Fig. 1b.  It is seen in this figure, that although the volume fraction
of ``dangling'' bonds is rather small (roughly $0.1$), it is constant,
indicating that the infinite-cluster probability undergoes a first-order
jump at the rigidity threshold.  The stressed backbone however shows a
non-trivial scaling, so that $n_B \sim L^{D_B}$, with $D_B = 1.78\pm
0.02$ as found previously[9,10].  Note that a fit to the
infinite-cluster data of Fig. 1b  suggests that the infinite cluster is
fractal, as claimed by Thorpe and Jacobs[10].  However, by separating
the dangling ends from the backbone bonds as shown in Fig. 1b, the
weakly first order character of $P_{\infty}$ is evident.  Thus in the
thermodynamic limit, we predict,
\begin{equation}
P_{\infty} = a + b (p-p_*)^{\beta'},
\end{equation}
where for the triangular lattice, the backbone exponent $\beta_t' =
0.255 \pm 0.03$.  Here we have used the result, $2 - \beta'/\nu = D_B$,
with $\nu =1.16\pm 0.03$ from our previous calculations[9,10].  For the
site-diluted triangular lattice the first-order jump in $P_{\infty}$,
$a_{sd}^t = 0.086\pm 0.005$, while for the case of bond dilution
$a_{bd}^t = 0.11\pm 0.02$ (see Fig. 1b).
\begin{figure}[] \vbox{ 
\centerline{ 
{\bf a} \psfig{figure=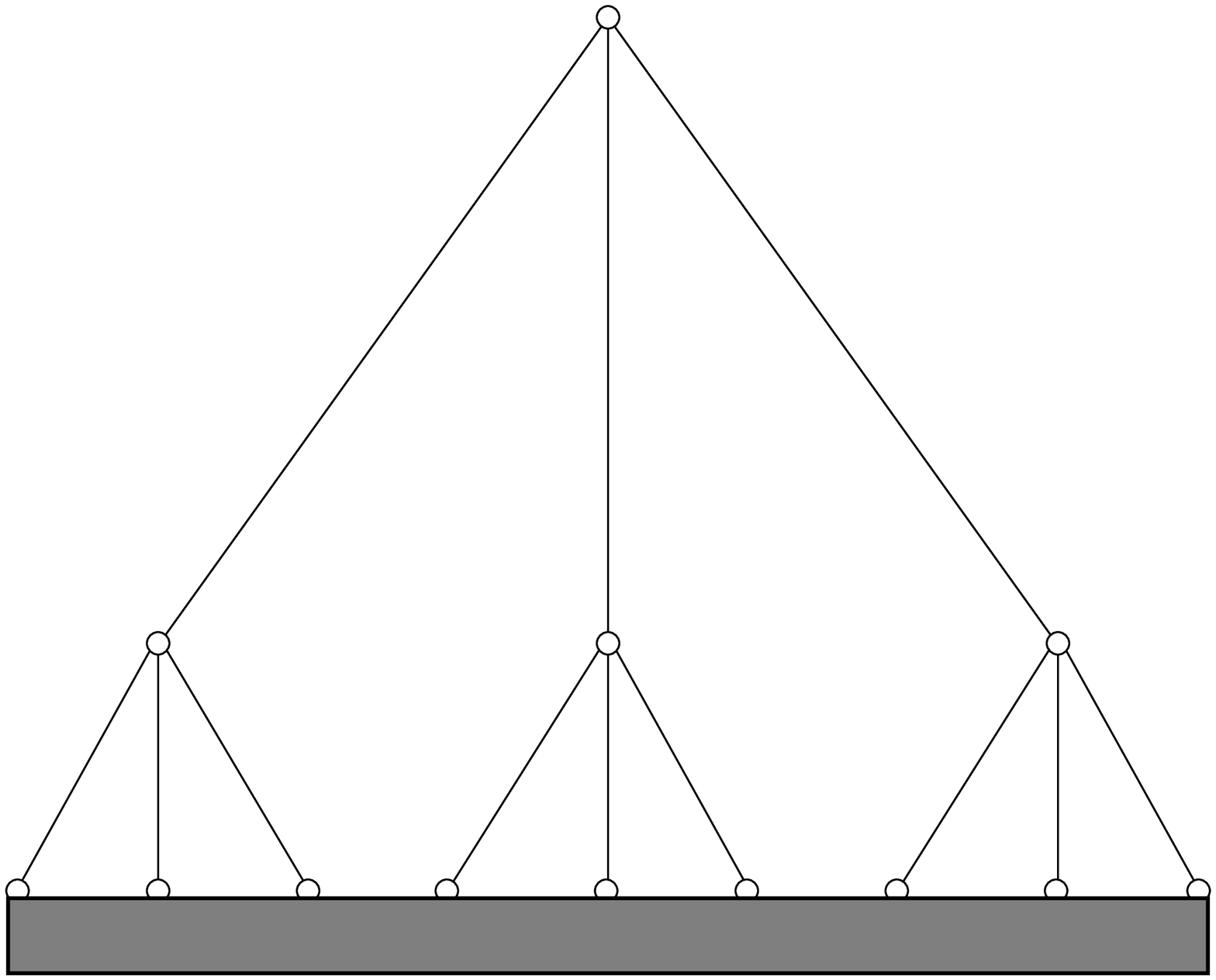,width=7cm,angle=270} \hfill
{\bf b} \psfig{figure=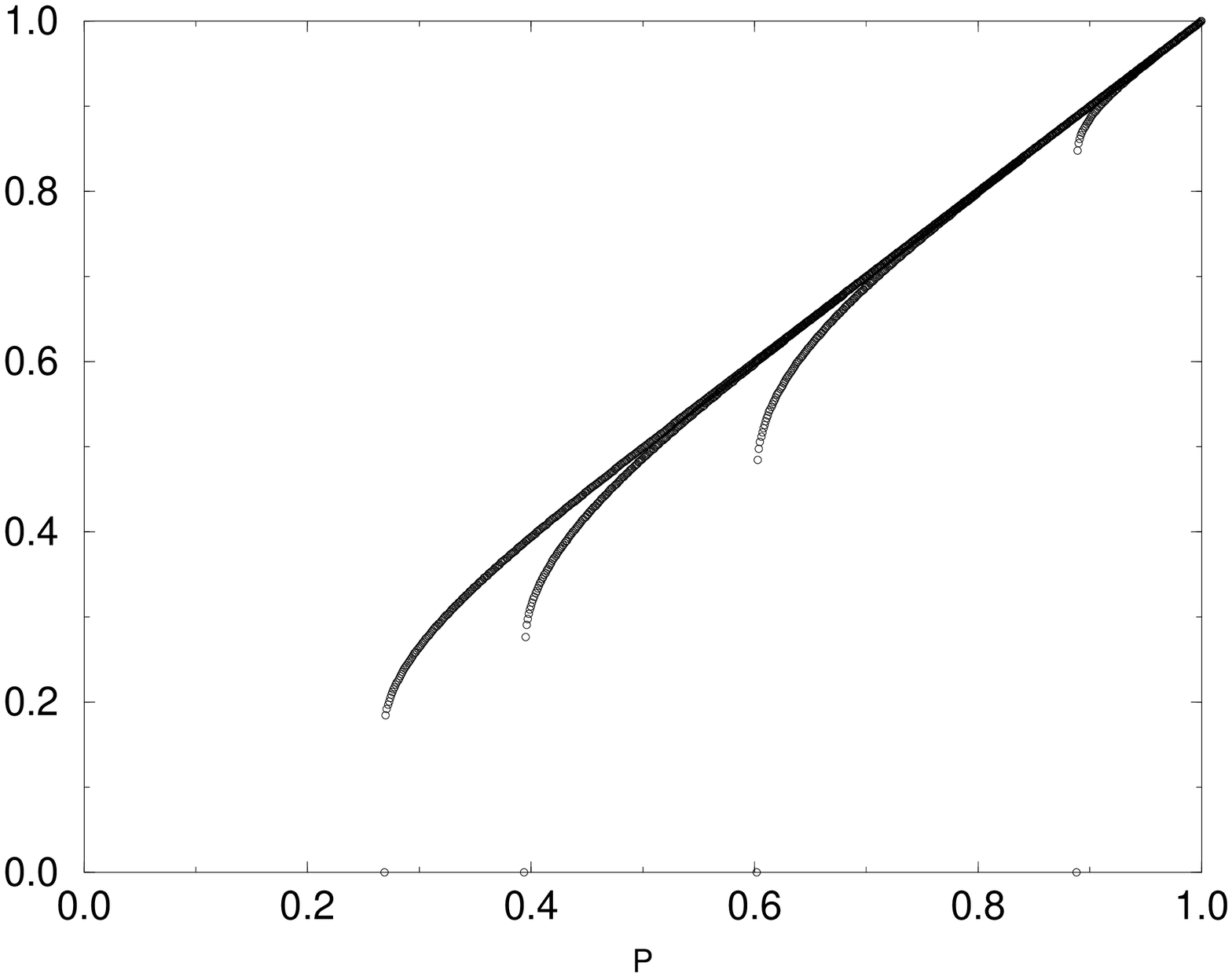,width=7cm,angle=270} }
\centerline{}
\caption{ 
{\bf a)} One branch of a $z=4$ Cayley tree.  The
lowest-level branches of the tree are attached to a rigid border. {\bf
b)}  $P_{\infty}$ for site-diluted trees with $g=2$ and, starting with
the rightmost curve, $z=4, 6, 9, 13$.
}
\label{fig:two}
} \end{figure}
Our mean-field theory uses exact ``constraint counting''(see below) on
trees.  We assign each node of a lattice $g$ degrees of freedom and
co-ordination $z$.  On the triangular lattice treated in the previous
paragraph, each node has two degrees of freedom and $z=6$ for the pure
lattice.  If the nodes of a triangular lattice were {\it extended
objects} or {\it bodies} instead of being pointlike ``joints'', then
each node would have an additional rotational degree of freedom.  Cases
of physical interest are then:
\begin{equation}
g=1 \ \  for\ connectivity \ percolation
\end{equation}
\begin{equation}
g=d\ \  for\  a\  joint
\end{equation}
\begin{equation}
g=d + {d(d-1) \over 2}\ \ for\ a \ body,\\
\end{equation}
where $d$ is the dimensionality of the embeding space.  Knowledge of $g$
and $z$ is enough to write down the simplest constraint-counting
mean-field theory[3].  Consider bond dilution with probability $p$ that
a bond is present. Present bonds restrict the motion of the nodes, and
hence they impose ``constraints'' on the degrees-of-freedom of the
lattice.   We assume that all of the constraints are ``independent'' (a
``dependent'' bond can be removed without causing any reduction in the
size of the rigid clusters), then the number of degrees of freedom that
remain unconstrained (the ``floppy'' modes per site $f$) is approximated
by,
\begin{equation}
f = g - pz/2  \ \  for \ \ p< p_* = {2g \over z}.
\end{equation}
For $p>p_*$, $f=0$ in this approximation.   This mean-field theory has
been useful in the study of glasses[3], but has yielded little
information about $P_{\infty}$, $P_B$ and $P_D$.  More recently, a
continuum mean field theory has been developed[11], which focuses on
$P_{\infty}$.  That theory indicates that the infinite-cluster
probability undergoes a first-order transition at the rigidity
threshold, but the connection with the key parameters $g$ and $z$ are
unclear.  In addition, in that work a pathological lattice model(a
 square lattice with random diagonals) was suggested as a paragidgm for
the rigidity transition. We show later why that model is anomalous.
\\
We now present our Cayley-tree model for rigidity percolation, which
provides a complete mean-field model for arbitrary $g$ and $z$.  The
trees have co-ordination number $z$ but, as usual in tree models, the
key results come from consideration of one branch of a tree (see Fig.
2a).  Consider site-diluted trees which are grown from a rigid boundary
at infinity.  Building inward from this boundary, we keep track of the
number of degrees of freedom of a node with respect to the boundary.
Rigidity can only be transmitted to higher levels of the tree if there
are enough rigid bonds present to offset the $g$ of degrees of freedom
of a newly-added node.  For connectivity percolation only one bond is
needed.  If a node is added to a $g=2$ tree, two bonds are needed to
offset the two degrees of freedom of the added node.  In general, if the
nodes of the tree have $g$ degrees of freedom, rigidity is transmitted
 to the next level of the tree provided the node is occupied, and
provided at least $g$ of the lower-level nodes are rigid.  This gives
the recurrence relation,
\begin{equation}
T^{l+1}_{\infty}=p \sum_{k=g}^{z-1} {z-1 \choose k}
(T^l_{\infty})^k(1-T^l_{\infty})^{z-1-k}
\end{equation}
Where $T^l_{\infty}$ is the probability that a site which is $l$ levels
from the rigid border (which is level 0) is rigid with respect to the
border.  For $g=1$ this reproduces the familiar model for connectivity
percolation[14], while for $g>1$ it is equivalent to so-called
``bootstrap'' percolation[15] (Note that the equivalence of the
transmission of rigidity and bootstrap percolation does not apply to
regular lattices).  If we take the thermodynamic limit (very large $l$),
Eq. (6) iterates to a steady-state solution, which we call $T_{\infty}$.
Finally we find $P_{\infty}$ from
\begin{equation}
P_{\infty} = p \sum_{k=g}^z {z \choose
k}(T_{\infty})^k(1-T_{\infty})^{z-k},
\end{equation}
The results for $T_{\infty}$ for several values of $g$ and $z$ are
presented in Fig. 2b.  It is seen from this Figure that for $g=2$, the
rigidity transition is first order, we show elsewhere that this
conculsion holds for all $g>1$[16].  An exact solution is
straightforward for the case $z=4$, $g=2$, where (from Eq. (6))
\begin{equation}
T_{\infty} = p(T_{\infty}^3 + 3T_{\infty}^2(1-T_{\infty})),
\end{equation}
which has the trivial solution $T_{\infty}=0$ and the non-trivial
solutions
\begin{equation}
T_{\infty} = { 3 \pm \sqrt(9-8/p) \over 4}.
\end{equation}
In order to ensure that $T_{\infty}=1$, when $p=1$, take the positive
solution in Eq. (9).  The square root in Eq. (9) becomes imaginary at
$p_* = 8/9$.  As $p\rightarrow p_*$, Eq. (9) yields,
\begin{equation}
T_{\infty} = 3/4 + c (p-p_*)^{1/2}
\end{equation}
The key feature of this equation is that $T_{\infty}$ is first order
{\it and} there is a {\it square-root singularity} superimposed upon the
first-order transition.  We find a similar behavior for all $g$ and $z$,
and so we write in general,
\begin{equation}
P_{\infty} = a + b (p - p_*)^{1/2},
\end{equation}
where $a$, $b$ and $p_*$ are dependent on $g$ and $z$.  Notice that this
is of the same form as Eq. (1) for the triangular lattice, and that the
square-root behavior is due to the backbone.  We thus argue that in
general the form Eq. (1) is correct, with the mean-field
backbone-exponent $\beta'_{mf} = 1/2$.   Full details of the Cayley-tree
result will be published elsewhere[16].
\\
 Due to the fractal backbone, there is a diverging length, and the
scaling properties are controlled by this diverging length.  As shown in
previous calculations this ``correlation'' length diverges with exponent
$\nu = 1.16 \pm 0.03$ [9,10] at the rigidity threshold in two
dimensions.  Coniglio's relation then applies to this problem[12], so
that we expect $n_r$, the number of red or ``cutting'' bonds to diverge
as $L^{1/\nu}$ at the rigidity threshold.  We find that to be the case
for both the site- and bond- diluted triangular lattices (see Fig. 3),
from which we find $P_r \sim n_r/L^2 \sim L^{x-2}$, with $x = 0.86 \pm
0.02 = 1/\nu$ confirming our previous result[9].
\begin{figure}[] \vbox{ 
\centerline{ \psfig{figure=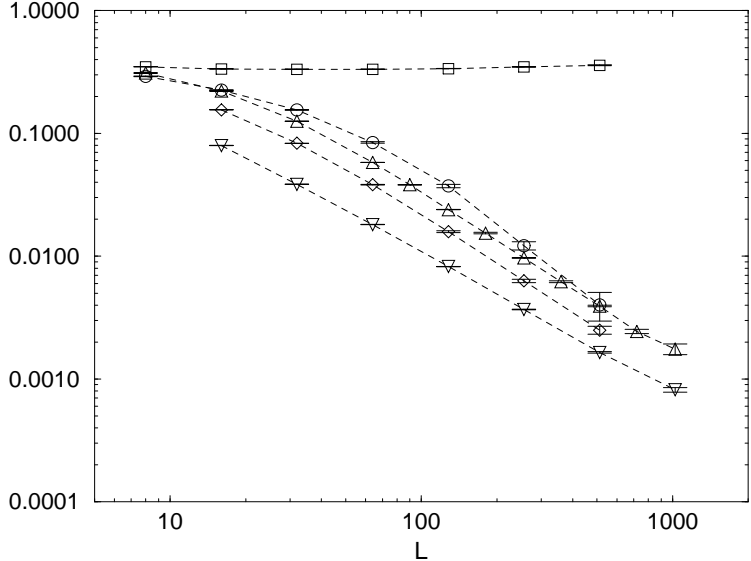,width=10cm,angle=270} }
\centerline{}
\caption{ 
The volume fraction of red bonds $P_r$ at the rigidity
threshold for: The triangular lattice with site dilution (inverted
triangles) and bond dilution(triangles); The ``random diagonal'' model
(see text) with $q=0$ (squares-this is the anomalous model of Fig. 4),
$q=0.1$(circles) and $q=0.4$(diamonds).
}
\label{fig:three}
} \end{figure}
Finally, we analyse and extend a simple model for gels, which has been
developed by Obukov[11] to provide an illustration of a first-order
rigidity transition (see Fig. 4 - the boundary conditions do not change
the conclusions).  The random diagonals are present with probability
$p_d$, and  all that is required to make the lattice rigid is one
diagonal present in every row of the square lattice.  The probability
that an infinite-cluster exists is then $P_+ = (1-(1-p_d)^L)^L$, from
which we find $p_{d*} \sim lnL/L$.  However, the resulting infinite
cluster contains the whole lattice, the rigid backbone is extensive, and
so is the number of cutting bonds (see Fig. 3 for a calculation of the
cutting bonds).  This is inconsistent with the result Eq. (1) and also
with the presence of a diverging length at the rigidity transition.
However, the model of Fig. 4 is atypical, as can be seen by considering
a generalized model in which we randomly add the diagonals (with
probability $p_d$) to a square lattice whose bonds have been diluted
with probability $q$ Obukov's model is $q=0$), while  if $q=1-p_d$ this
model is equivalent to the bond-diluted triangular lattice.  We find
that even for a small amount of dilution, e.g.  $q=0.10$, the rigidity
transition returns to the behavior characteristic of the isotropic
triangular case (see Fig. 3).  We find that for sufficiently large
lattice sizes, the universal behavior Eq.  (1) holds for any finite
$q<0.5$, and that the ``fully-first-order'' transition (i.e. a
first-order backbone) only occurs in the special case of a perfect
square lattice(or other ``marginal'' regular lattices) with randomly
added diagonals.  We have found similar pathological cases on Cayley
trees.
\begin{figure}[] \vbox{ 
\centerline{ \psfig{figure=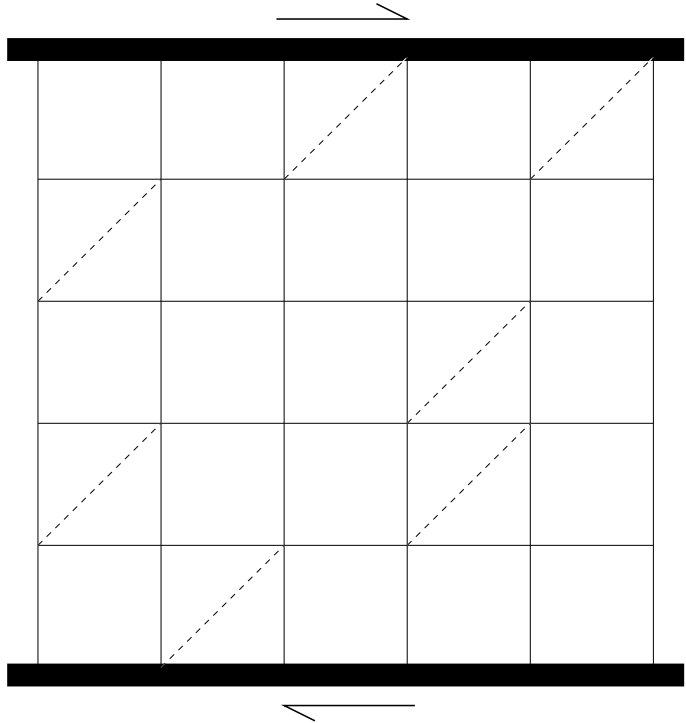,width=9cm,angle=270} }
\centerline{}
\caption{ 
The square lattice with random diagonals and no removed
horizontal or vertical bonds, so $q=0$ (see text).
}
\label{fig:four}
} \end{figure}
\\
PMD acknowledges the US Department of Energy under contract
DE-FG02-90ER45418 for support, Mike Thorpe and Don Jacobs for useful
discussions. CM acknowledges support from the Conselho Nacional de
Pesquisa CNPq, Brazil.\\
\\
\newpage

\noindent {\bf References}\\

\noindent
 [1] D. Stauffer and A. Aharoni ``Introduction to percolation theory''
 second edition,  Taylor and Francis (London 1992)\hfill\break
 [2] E. Guyon, S. Roux, A. Hansen, D. Bideau, J.-P. Troadec and H.
 Crapo, Rep. Prog. Phys. {\bf 53}, 373 (1990)\hfill\break
 [3] M.F. Thorpe, J. Non-Cryst. Sol. {\bf 57}, 355 (1983)\hfill\break
 [4] M. Rubinstein, L. Leibler and J. Bastide, Phys. Rev. Lett. {\bf
 68}, 405 (1992)\hfill\break
 [5] S. Feng and P.N. Sen, Phys. Rev. Lett. {\bf 52}, 306
 (1984)\hfill\break
 [6] A.R. Day, R.R. Tremblay and A.-M.S. Tremblay, Phys. Rev. Lett. {\bf
 56}, 2501 (1986)\hfill\break
 [7] A. Hansen and S. Roux, Phys. Rev. {\bf B40}, 749 (1989)\hfill\break
 [8] S. Arbabi and M. Sahimi, Phys. Rev. {\bf B47}, 695
 (1993)\hfill\break
 [9] C. Moukarzel and P.M. Duxbury, Phys. Rev. Lett. {\bf 75}, 4055
 (1995); (see cond-mat/9612237). \hfill\break
 [10]  D. Jacobs and M.F. Thorpe, Phys. Rev. Lett. {\bf 75}, 4051
 (1995);
\\
D. Jacobs and M.F. Thorpe, Phys. Rev. {\bf E53}, 3682 (1996)\hfill\break
 [11] S.P. Obukhov, Phys. Rev. Lett. {\bf 74}, 4472 (1995)\hfill\break
 [12] A. Coniglio, J. Phys. {\bf A15}, 3829 (1982);
\\
 C.~Moukarzel, to be published. \hfill\break
 [13] Bruce Hendrickson, Siam J.~ Comput.{\bf 21},65 (1992);
\\
C. Moukarzel, J.~Phys.~A: Math.~Gen. {\bf 29}, 8097 (1996) (see
physics/9612013).\hfill\break
 [14]  M.E. Fisher and J.W. Essam, J. Math. Phys. {\bf 2}, 609
 (1961)\hfill\break
 [15]  J. Chalupa, P.L. Leath and G.R. Reich, J. Phys. {\bf C12}, L31
 (1979);
\\
C.M. Chaves and B. Koiller, Physica {\bf A218}, 271 (1995)\hfill\break
 [16]  C. Moukarzel, P.M. Duxbury and P.L. Leath, to be
 published.\hfill\break
 \\

\end{document}